\documentclass[aoas,MSNbibl,nameyear,dvips]{arximspdf}
\usepackage{graphicx}

\doi{10.1214/14-AOAS749} 
\volume{8}
\issue{3}
\pubyear{2014}
\firstpage{1671}
\lastpage{1689}
\docsubty{FLA}



\begin{document}
\begin{frontmatter}

\title{Degradation-based residual life prediction under different environments}
\runtitle{Degradation modeling under varying environments}

\begin{aug}
\author[A]{\fnms{Rensheng}~\snm{Zhou}},
\author[A]{\fnms{Nicoleta}~\snm{Serban}\corref{}\ead[label=e2]{nserban@isye.gatech.edu}}
\and
\author[A]{\fnms{Nagi}~\snm{Gebraeel}}
\runauthor{R. Zhou, N. Serban and N. Gebraeel}
\affiliation{Georgia Institute of Technology}
\address[A]{H. Milton Stewart School of Industrial\\
\quad and Systems Engineering\\
Georgia Institute of Technology\\
Atlanta, Georgia 30331\\
USA\\
\printead{e2}} 
\end{aug}
%

\received{\smonth{7} \syear{2013}}
\revised{\smonth{4} \syear{2014}}

%
\begin{abstract}
Degradation modeling has traditionally relied on historical signals to
estimate the behavior of
the underlying degradation process. Many models assume that these
historical signals are acquired under the same environmental conditions
and can be observed along the entire lifespan of a component. In this
paper, we relax these assumptions and present a more general
statistical framework for modeling degradation signals that may have
been collected under different types of environmental conditions. In
addition, we consider applications where the historical signals are not
necessarily observed continuously, that is, historical signals are
sparse or fragmented. We consider the case where historical degradation
signals are collected under known environmental states and another case
where the environmental conditions are unknown during the acquisition
of these historical data. For the first case, we use a classification
algorithm to identify the environmental state of the units operating in
the field. In the second case, a clustering step is required for
clustering the historical degradation signals. The proposed model can
provide accurate predictions of the lifetime or residual life
distributions of engineering components that are still operated in the
field. This is demonstrated by using simulated degradation signals as
well as vibration-based degradation signals acquired from a rotating
machinery setup.
\end{abstract}

%
\end{frontmatter}

\section{Introduction}

\textit{Degradation signals} are signals that are correlated with
physical degradation processes that take place prior to failures of
engineering systems or components. For this reason, degradation signals
are commonly used as indicators of the health status or the performance
level of functioning components. In degradation data analysis, an
engineering component is considered to have failed once its degradation
level reaches a fixed and prespecified critical level, known as the
\textit{failure threshold}.

\begin{figure}[b]

\includegraphics{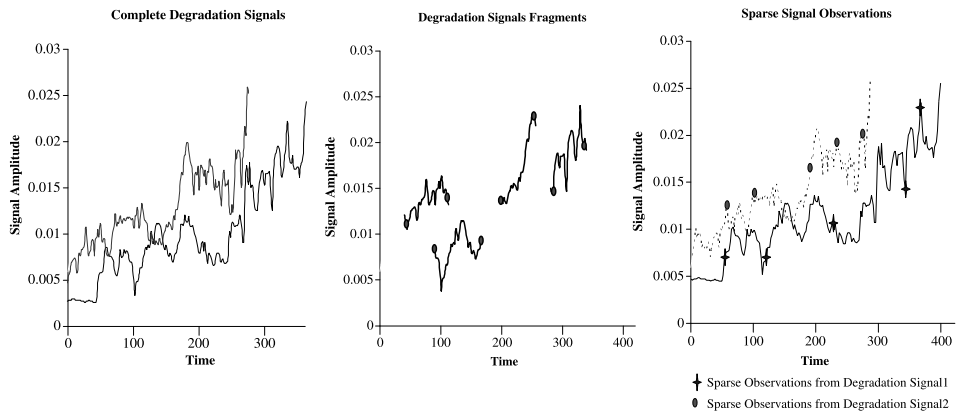}

\caption{Examples of complete, fragmented and sparse degradation signals.}\label{figexsignals}
\end{figure}

Recent developments in degradation modeling, such as \citet{geb05}, \citet{Liao06b}, \citet{zhou2} and \citet{zhou1},
have focused on utilizing degradation-based signals to predict lifetime
or residual life distributions of engineering components. Almost all of
the existing models rely on a historical database of degradation
signals for estimating model parameters specifying the behavior of the
degradation process. These signals can be acquired through a variety of
methods. For instance, it may be possible to acquire frequent
observations through extensive monitoring of a component over its life
span, which results in a \textit{complete degradation signal}. Another
alternative is to follow an intermittent monitoring strategy, which
leads to \textit{incomplete degradation signals}. For instance, the
degradation signals could be sparsely observed (i.e., sparse
degradation signals) or densely observed over short time intervals
(i.e., fragmented degradation signals). An example of complete, sparse
and fragmented degradation signals is provided in Figure~\ref
{figexsignals} available from \citet{zhou2}.

More generally, components may be operated under different
environmental conditions, for instance, different levels of humidity,
speeds, loads and temperatures, among others. Environmental conditions
can significantly accelerate or decelerate the degradation processes of
functioning components. For example, in \citet{geb08},
bearings are run at different rotating rates and, as a result, these
bearings degrade at significantly different rates. However, most
existing literature on degradation modeling assumes that components are
from the same population and are operated under the same environmental
conditions. An approach that does take the environmental conditions
into account is commonly used for modeling accelerated degradation test
(ADT) data. \citet{Whitmore97} propose a Wiener diffusion
process with a time-scale transformation that depends upon the level of
stress under which the ADT signals data are observed. Similar ideas can
be found in \citet{Doksum92}, \citet{Liao06}, \citet{park06} and \citet{Tseng09}. Other approaches that
incorporate environmental data in degradation models include \citet{Kharoufeh03}, \citet{Kharoufeh05}, \citet{Li05} and \citet{Singpurwalla95}. One common characteristic among these existing methods is that
the dependence of the degradation processes on the environmental
conditions is specified using a parametric functional form that
describes how the degradation processes evolve over time; see \citet{Bae04}, \citet{geb06}, \citet{lawless04}, \citet{Meeker98}, \citet{Robinson00},
\citet{Wang10b} and \citet{Whitmore98}, among others.

In some applications, the underlying physics of degradation processes
indeed may be known in advance. However, in many applications it may be
difficult to identify a parametric model that can accurately capture
the underlying trend of degradation processes. To overcome this
challenge, recent research has considered nonparametric degradation
models, in which the functional form is learned from the degradation
data. \citet{Shiau99} applied nonparametric regression techniques
to characterization of degradation signals of a light emitting diode
product under different stress levels. \citet{muller05}
proposed a time-varying regression approach for predicting the
remaining lifetime of flies based on the observed reproductive
activity. Both research works, however, assume that the degradation
signals are completely observed. In \citet{Liao11}, \citet{zhou2} and \citet{zhou1}, the authors pointed out that the challenge is even more
noteworthy when there are only incomplete degradation signals
available. To overcome this challenge, they developed nonparametric
degradation models based on functional data analysis and demonstrated
that these models are generally more flexible and more robust to model
misidentifications. These nonparametric models apply to signals
observed at a small number of nonregularly sampled time points given
that the number of degradation signals is sufficiently large.

In this paper, we develop a nonparametric model that does not require
the functional form of mean degradation trend to be known in advance.
This allows for more flexibility in modeling as compared to parametric
approaches such as \citet{meeker93} and other references mentioned
above. More specifically, we build our modeling framework based upon
functional data analysis (FDA) techniques. Functional data analysis is
a collection of statistical techniques that focus on analyzing data in
the form of curves, surfaces or functions. Examples of FDA
methodologies include functional principal component analysis in \citet{Yao05},
functional regression analysis in \citet{Ramsay91}, functional time warping analysis in \citet{Telesca08} etc.
A comprehensive review and discussions of FDA methods and
applications can be found in \citet{Ramsay05}. In this paper,
we apply FDA techniques to different types of degradation signals,
whether they are complete or incomplete, that specify the effects
of varying environmental conditions on degradation processes within a
flexible framework not requiring prior knowledge about the behavior of
the degradation trend. Using a similar Bayesian framework as in \citet{zhou2}, \citet{zhou1} but implemented within the model specifications in
this paper, we predict and update in real time the residual life
distributions (RLD) of components operated in the field, referred to as
\textit{fielded components}, by using their partially observed degradation
signals. Unlike previous work by \citet{zhou3}, the model
presented in this paper allows for the degradation signals to be
observed under varying environment conditions.

Specifically, we assume that the environmental conditions can be
categorized into discrete types, and they are time invariant. Under
this assumption, we consider two different scenarios. One scenario is
supervised learning, in which the environmental types for all the
training signals are available. In this case, only the new test
(fielded) component's environmental type is unknown and needs to be
predicted. Another scenario is unsupervised learning, in which the
environmental information is not available in advance. Under the second
scenario, we also need to learn the clustering of the different
environments along with the estimation of the degradation process
corresponding to each group of environments. In both scenarios, the
degradation model includes a random variable describing the cluster
membership or the type of environmental conditions. In the second
scenario, this variable is latent or missing.

Because we have two sources of missing data, one due to the fact the
signals are thresholded and the second due to the missing cluster
membership or unknown environment type, we propose using an
Expectation--Maximization algorithm to estimate and update the
distribution of the degradation process. The EM-type algorithm is a
more flexible approach to model estimation when signals come from
different environments.

The performance of the developed degradation framework is demonstrated
by using both simulated degradation signals and a case study from a
rotating machinery setup. We consider extensive types of scenarios, for
instance, the components may be operated under known or unknown types
of environmental conditions; the degradation signals may be complete or
incomplete; the underlying degradation trend may or may not be expanded
from the basis functions we specify. The results indicate that the
proposed framework is quite flexible and can accurately predict the RLD
of components operated in the field under all these scenarios.

The remaining paper is organized as follows. We first discuss the
general model in Section~\ref{model}. We present the details of the
estimation approach in Section~\ref{modelest} followed by model
prediction in Section~\ref{modelpred}. To assess the performance of our
methodology, we continue with a simulation study and a case study in
Sections~\ref{study}~and~\ref{study2}, respectively. Conclusions
and some discussions are given in Section~\ref{discuss}. Technical
details are provided in the supplemental material [\citet{ZSG14}].

\section{The model}\label{model}

\subsection{Model decomposition}

Denote the degradation level at time $t$ by $S(t)$, the failure
threshold by $D$ and the environmental type by $Z$. According to the
definition of failure, the lifetime of a component, denoted by $T$, is
%
\begin{equation}
\label{eqlifetime} T = \inf_{t} \bigl\{S(t)\geq D \bigr\}.
\end{equation}
Here, $T$ is random and not observable due to the incompleteness and
discreteness of the observed degradation signals.

In this paper, we consider a nonparametric decomposition of $S(\cdot)$,
by assuming that it can be represented by a set of basis functions
$B(\cdot)$, with a vector coefficient denoted by $\gamma$. We also
assume that the environmental conditions are specified by the variable
$Z$. Based on these assumptions, we consider estimating the degradation
model using a likelihood based approach. The likelihood decomposition
used in our model estimation is motivated by the fact that $T$ is
unobservable. The decomposition is
\[
{L}(S,Z,\gamma) = f(Z)f(\gamma|Z)f(S|Z,\gamma), %
\]
where $Z$ and $\gamma$ are latent variables and, thus, we need to
impose a parametric structure on both $Z$ and $\gamma|Z$. It is natural
to assume that $Z$ follows a multinomial distribution, as the
contribution of each environment type will be given by the
probabilities of the multinomial distribution. This assumption in turn
specifies that $S(\cdot)$ is a mixture process. Furthermore, the
proportional parameters of the multinomial distribution can be
estimated by the fraction of each cluster in the historical data set or
determined by prior knowledge.

The distribution of $f(\gamma|Z)$ can be approximated by, for instance,
a Gaussian distribution. This implies that, unconditionally, $\gamma$
follows a mixture of Gaussian distributions. The number of mixtures is
equal to the number of different values $Z$ can take corresponding to
the number of different environments. Other parametric assumptions can
be considered at the price of a higher computation cost.

The second step is to specify $f(S|Z,\gamma)$. In our model
specification in Section~\ref{modeldeg}, the observed degradation
signal is a sum of the underlying degradation process, which is
completely determined by the basis coefficient $\gamma$ and a
measurement error term. Therefore, $f(S|Z,\gamma)$ is fully determined
by the distributional assumption about the measurement error. For ease
of computational efficiency and for a close form expression of the RLD
predictions, we assume the error term to follow a Gaussian distribution.

Given these specifications, we can predict the residual life of the
component operated in the field in two steps. The first step is to
predict $f(\gamma|S)$, the posterior distribution of $\gamma$ given the
partial observations of the component still operated in the field. At
the second step, the RLD of the operated component can then be
predicted following the definition of the failure time in (\ref{eqlifetime}).

\subsection{Modeling degradation signals}\label{modeldeg}
Denote the measurement (or inspection) time by $t_{lj}$, where $l = 1,
\ldots, L$ ($L$ is the number of signals or components) and $j = 1,
\ldots,n_l$ ($n_l$ is the number of observation time points for
component $l$). We assume that the time points are prespecified within
a bounded interval $[0,M]$, where $M$ refers to the maximum
experimental time. The degradation amplitudes of the component $l$ are
denoted by $S_l=(S_l(t_{lj}),\ldots,S_l(t_{ln_l}))$.

Note that $S_l(t_{lj})$ may not always be observable. For instance, a
component may be shut down or replaced instantaneously after its
degradation level reaches the failure threshold. In other words, no
further observations can be acquired beyond the failure threshold.
These types of signals are referred to as \textit{truncated degradation
signals} in \citet{zhou2}. In these applications, $S_l(t_{lj})$ is
observable only if the component $l$ has not failed by time $t_{lj}$.

We assume that the underlying degradation process, denoted by $X(\cdot
)$, can be represented by a fixed number of basis functions. Based on
this assumption, we consider the following degradation model specifying
the conditional distribution $f(S|Z,\gamma)$:
%
\begin{equation}
\label{model1} S_{l}(t)=X_{l}(t)+\varepsilon_{l}(t)=B(t)
\gamma_{l}+\varepsilon_{l}(t),
\end{equation}
%
where:

\begin{itemize}
\item$X_{l}(\cdot)$ represents the underlying degradation process.
\item$B(\cdot)$ represents the basis functions of dimension $q$,
defined over the time interval $[0,M]$. For illustrative purposes, we
use the cubic B-spline bases because of its flexibility. A B-spline
function is a function that is connected by polynomial pieces with
specified orders (``cubic'' corresponds to the order 4). Cubic B-spline
bases have been widely used in the literature for modeling smooth
functions [\citet{Eilers96}].
\item$\gamma_{l}$ represents the basis coefficient for the $l$th
signal. It is a vector of dimension~z$q$.
\item$\varepsilon(\cdot)$ represents the error term. We assume that
$\varepsilon(\cdot)$ is independent and identically distributed at
different time points.
\end{itemize}

\subsection{Modeling environmental clusters}\label{modelclassify}

In this paper we assume a component's environmental type is time
invariant, that is, it does not change over time. Let the environmental
type for component $l$ be $Z_{l}\in\{1,2,\ldots,K\}$, where $K$
represents the number of environmental types or clusters (in the
remaining paper, ``environmental types'' and ``clusters'' will be used
interchangeably). We make the following distributional assumptions:

\begin{itemize}
\item$Z_l$ follows a multinomial distribution with parameters $\pi=
(\pi_1,\ldots,\pi_K)$.
\item Conditional on $Z_l$, the basis coefficient $\gamma$ follows a
normal distribution. The distributional means and variances are
different among environmental types. More specifically, $\gamma
_{l,k}\equiv\gamma_{l}|Z_l=k \sim N(\mu_k, \Lambda_k)$, where ${\bolds\mu
}_k = ({\bolds\mu}_{k1},\ldots,{\bolds\mu}_{kq})^T$ and $\Lambda_k$ is a
$q\times q$ matrix.
\item Conditional on $Z_l$, the error terms are assumed to follow a
normal distribution. The variances are different across clusters. In
other words, $\varepsilon_l(t)|Z_l=k \sim N(0, \sigma_k^2)$.
\end{itemize}

In summary, we have the following model:
%
\begin{equation}
\label{model2} \cases{ \displaystyle Z_{l} \sim\operatorname{Multinomial}(
\pi_1,\ldots,\pi_K), \vspace*{3pt}
\cr
\displaystyle
\gamma_{lk} \equiv\gamma_{l}|(Z_{l}=k)\sim N(
\mu_k, \Lambda_k), \vspace*{3pt}
\cr
\displaystyle
S_{l}(t) =B(t) \gamma_{l}+\varepsilon_l(t),
\vspace*{3pt}
\cr
\displaystyle \varepsilon_l(t)|(Z_{l}=k)
\sim N \bigl(0, \sigma_k^2 \bigr).}
\end{equation}

Based on the above formulation, we have $S_{l}(t)|Z_{l}=k \sim N(B(t)\mu
_k$,\break $B(t)\Lambda_k B(t)^T+\sigma_k^2I)$.

\section{Estimation}\label{modelest}

As mentioned earlier, we will consider two possible scenarios, that is,
the cluster membership for the training components may or may not be
known a priori. In the machine learning context, this corresponds to
classification and clustering problems, respectively.

Let $\mu=(\mu_1,\ldots,\mu_K)$, $\Lambda=(\Lambda_1,\ldots,\Lambda_K)$,
$\pi=(\pi_1,\ldots,\pi_K)$, $\sigma=(\sigma_1,\ldots, \sigma_K)$. The
vector $\theta=(\mu,\Lambda,\pi,\sigma)$ includes all the parameters of
the model in (\ref{model2}). Because of the presence of latent
variables, it is intractable to maximize the complete data
log-likelihood directly with respect to these parameters. To address
this challenge, we apply an EM algorithm in order to obtain the maximum
likelihood estimate of $\theta$. The estimation procedures are similar
for the classification and clustering scenarios, except for an extra
step in the clustering case, in which we classify all the training
units. Details about the estimation algorithm are provided in the
supplemental material [\citet{ZSG14}]. In the following subsections, we
highlight the challenge of estimating the covariance matrix $\Lambda$
and discuss how to determine the tuning parameters.

\subsection{Estimating the covariance matrix}

To allow for more flexibility, we assume that $\Lambda_k$ are different
across clusters. This implies that we need to estimate $\frac{{Kq(q +
1)}}{2}$ parameters for the covariance matrix. If we do not have a
sufficiently large historical data set for training, then the
covariance matrix estimate will be unstable and inaccurate. To overcome
this challenge, we follow the idea of regularized discriminant analysis
(RDA) proposed in \citet{Friedman89}. More specifically, we regularize the
raw covariance matrix estimates in two steps:

\begin{longlist}[\textit{Step} 2.]
\item[\textit{Step} 1.] Shrink the individual sample covariance matrix
estimate (${\hat\Lambda_k}$) toward the population sample covariance
matrix estimate (${\hat\Lambda}$) with a parameter $0 \le\lambda \le1$:
\[
{{\hat\Lambda}_k}(\lambda) = (1 - \lambda){{\hat
\Lambda}_k} + \lambda\hat\Lambda. %
\]

\item[\textit{Step} 2.] Shrink ${\hat\Lambda_k}(\lambda)$ toward a multiple
of the identity matrix with a parameter $0\le\zeta \le1$:
\[
{{\hat\Lambda}_k}(\lambda,\zeta) = (1 - \zeta){{\hat\Lambda
}_k}(\lambda) + \zeta\frac{{\operatorname{tr}({{\hat\Lambda}_k}(\lambda))}}{p}I.
\]

\citet{Friedman89} demonstrates through numerous case studies that ${{\hat
\Lambda}_k}(\lambda,\zeta)$ is generally more stable and more
accurate than the raw covariance matrix estimate~${\hat\Lambda_k}$,
especially when the sample size of certain clusters is not large enough.
\end{longlist}

\subsection{Choice of tuning parameters}

The degradation model presented\break above depends on a series of tuning
parameters: the basis dimension $q$, the shrinkage parameters $\lambda$
and $\zeta$, and possibly the number of clusters $K$ (for the
clustering scenario). With larger values of $q$ and $K$, we have more
parameters to estimate, resulting in smaller estimation bias but higher
estimation
variance. Thus, we need to \mbox{select} these parameters in order to optimize
the bias-variance trade-off in the model.

To reduce the computational burden, we determine the turning parameters
in two steps. We first select $q$ and possibly $K$ from a set of
candidate values by following a cross-validation procedure.
Cross-validation is a model validation \mbox{technique} for assessing how
accurately a predictive model will perform in an independent data set.
A detailed explanation of the cross-validation procedures can be found
in \citet{Hastie05}.

In our context, we compute the RLD prediction error for each
combination of the candidate values in the cross-validation process.
Based on the error results, we choose the combination of $q$ and $K$
that yields the smallest error. In the second step, we follow a similar
cross-validation procedure to determine the optimal values of $\lambda$
and $\zeta$. The number of candidate values for these parameters
determines the computation time for finding the optimal values. It is
common to start with a rough set of candidate values that would provide
an approximate range for the optimal values and then refine it with
sample points within that range.

\section{Prediction}\label{modelpred}

Given the degradation signal $S^*$ of a new component operated in the
file observed up to time $t^*$, our goal is to predict its residual
life $\mbox{RL}^*$, that is, the time left for the signal to reach the failure
threshold $D$. In other words, we need to derive the density function
$f({\mbox{RL}^*}|{S^*})$. We approach this in two steps according to the
following equation:
\[
f\bigl({\mbox{RL}^*}|{S^*}\bigr) = \int_{{\gamma}} {f\bigl({
\mbox{RL}^*}|{\gamma},{S^*}\bigr)f\bigl({\gamma }|{S^*}\bigr)\,d{\gamma. }}
\]

\begin{longlist}[\textit{Step} 2.]
\item[\textit{Step} 1.] Compute ${f({\gamma}|{S^*}):}$
\[
f\bigl({\gamma}|{S^*}\bigr) = \sum
_{k = 1}^K {f
\bigl({\gamma}|{Z^*} = k,{S^*}\bigr)P\bigl({Z^*} = k|{S^*}\bigr).} %
\]

\begin{longlist}[(3)]
\item[(1)] ${{\gamma}|{Z^*} = k,{S^*}}$ follows a Gaussian distribution. Its
mean vector and covariance matrix can be computed based on the general
Bayesian linear theory. Details can be found in the supplemental material [\citet{ZSG14}].

\item[(2)] ${Z^*} = k|{S^*}$ follows a multinomial distribution with its
proportional parameters derived as follows. Since ${S^*}|{Z^*} = k \sim
N(B{\mu_k},B{\Lambda_k}{B^T} + {\sigma_k ^2}I)$, we have
%
\begin{eqnarray}\label{enviprob}
&& P \bigl({Z^*} = k|{S^*} \bigr)\nonumber
\\
&&\qquad =\frac{{f({S^*}|{Z^*} =
k){\pi_k}}}{{\sum_{j = 1}^K {f({S^*}|{Z^*} = j){\pi_j}} }}
\\
&&\qquad =\biggl(\bigl| B{\Lambda_k}{B^T} + \sigma_k^2I\bigr |^{ -1/2}\nonumber
\\
&&\hspace*{38pt}{}\times \exp \Biggl(  - \frac12\bigl({S^*} - B{\mu
_k}\bigr)^T\bigl(B{\Lambda_k}{B^T} + \sigma_k^2I\bigr)^{ - 1}\bigl({S^*} - B{\mu
_k}\bigr) \Biggr){\pi_k}\biggr)\nonumber
\\
&&\quad\qquad{}\Big/
\Biggl(\sum_{j = 1}^K \bigl| {B{\Lambda _j}{B^T} + \sigma_j^2I} \bigr|^{ - 1/2}\nonumber
\\
&&\hspace*{61pt}{}\times
\exp \biggl(  - \frac12{{\bigl({S^*} - B{\mu_j}\bigr)}^T}{{\bigl(B{\Lambda_j}{B^T} + \sigma_j^2I\bigr)}^{- 1}}\bigl({S^*} - B{\mu_j}\bigr) \biggr){\pi_j} \Biggr).\hspace*{-10pt}\nonumber
\end{eqnarray}
\end{longlist}

\item[\textit{Step} 2.] Compute $f({\mbox{RL}^*}|{S^*})$:
since $f({\mbox{RL}^*}|{S^*})$ does not have a closed-form expression, we
suggest using a parametric bootstrap [\citet{bootstrap2}] to
generate samples from $f({\mbox{RL}^*}|{S^*})$ as follows:

\begin{longlist}[(5)]
\item[(1)] Generate a random sample $\gamma$ from $f({\gamma}|{S^*})$
according to the density function given in step~1 (both ${\gamma
}|{Z^*} = k,{S^*}$ and ${Z^*} = k|{S^*}$ follow a well-defined
distribution that can be generated from existing statistical packages).
\item[(2)] Generate the corresponding signal $S_b\dvtx  S_b(t)=B(t)\gamma$.
\item[(3)] Get the residual life ${\mbox{RL}_b}$ for the generated signal
according to the failure time definition: ${\mbox{RL}_b} = \inf_{t} \{
{S_b}(t) > D\}-t^*$.
\item[(4)] If $\mbox{RL}_b>0$, then proceed to the next step; otherwise, repeat
the above steps until $\mbox{RL}_b>0$.
\item[(5)] Repeat the above steps for $N_b$ times and get $N_b$ values of
$\mbox{RL}^*\dvtx  \mathbf{RL}=(\mbox{RL}_{1},\mbox{RL}_{2},\ldots,\mbox{RL}_{N_b})$.
\end{longlist}

$\mathbf{RL}$ can then be used for the estimation of any statistics
related to $\mbox{RL}^*$, such as quantiles and prediction intervals.
\end{longlist}

\section{Simulation study}\label{study}

\subsection{Simulation setting}\label{study11}
In this study, we assume that components are from two different
clusters, that is, they are operated under two different environmental
types. We first simulate the cluster membership $Z_l$ from a Binomial
distribution with equal parameters, that is, $Z_l \sim \operatorname{Binomial}(\pi
_1=0.5,\pi_2=0.5)$. Next, we generate signals from each cluster based
on the following model settings:

\begin{itemize}
\item In\vspace*{1pt} cluster $1$, $S_{l}(t) = \mu(t)+ X_l(t)+\varepsilon_l(t)$, where:

\begin{itemize}
\item $\mu(t)= 4t^2e^{t/25}$, which represents the overall mean
degradation trend for components within this cluster.
\item $X_l(t)= \beta_lt^2$, which is introduced to account for the unit
to unit heterogeneity in degradation. Here, $\beta_l \sim N(0, 1.5^2)$.
\item$\varepsilon(t)|(Z_{l}=2) \sim N(0, 60^2)$.
\end{itemize}

\item In cluster $2$, $S_{l}(t) =B(t)\gamma_{l}+\varepsilon(t)$, where:

\begin{itemize}
\item$B(\cdot)$ represents the cubic B-spline basis with its dimension $q=5$.
\item$\gamma_{l} \sim N(\mu_1, \Sigma_1)$, where $\mu_1=(0, 500,
1500, 2500, 3000)$ and
\[
\Sigma_1 ^{ - 1} = {\Omega_1}/{5600 };\qquad {{
\Omega_1} } = \pmatrix{ 2 & { - 1} & 0 & 0 & 0
\cr
{ - 1} & 2 & { -
1} & 0 & 0
\cr
0 & { - 1} & 2 & { -1} & 0
\cr
0 & 0 & { - 1} & 2 & { - 1}
\cr
0 &
0 & 0 & { - 1} & 1 }_{5 \times5 }. %
\]
The value of $\mu_1$ ensures that the overall mean degradation trend of
this cluster is linear. The underlying degradation process of each
component can still be nonlinear. A covariance matrix of similar
structure to $\Sigma_1$ is frequently used as a prior for the basis
coefficients under the Bayesian framework [\citet{Lang04}].
Under the frequentist framework, this corresponds to penalized
regression splines [\citet{Eilers96}].
\item$\varepsilon(t)|(Z_{l}=1) \sim N(0, 80^2)$.
\end{itemize}
\end{itemize}
We note that the signals in cluster $2$ are generated under the general
framework proposed in equations (\ref{model2}), while the signals
within cluster $1$ are not.

\begin{figure}

\includegraphics{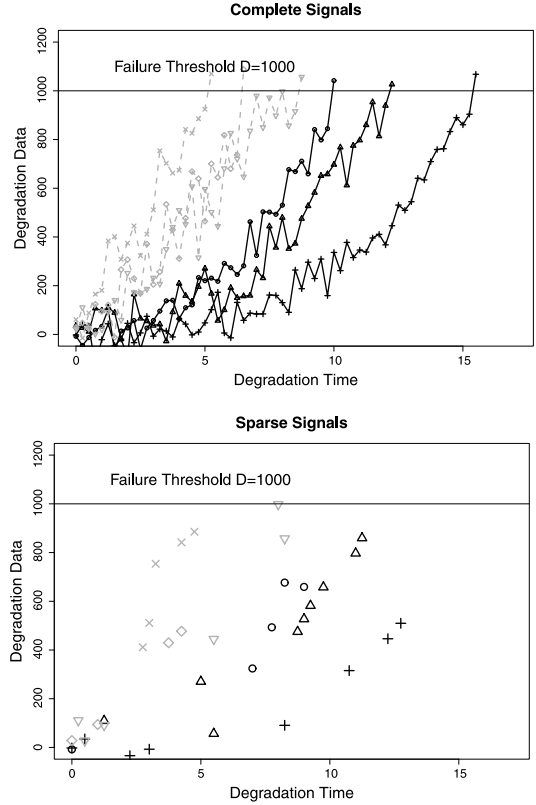}

\caption{Examples of complete (\textup{top plot}) and sparse (\textup{bottom plot})
degradation signals.}\label{figsimusignals}
\end{figure}

Based on the above model settings, we generate $100$ signals for
training the degradation model and another $100$ signals for evaluating
the RLD prediction performance of our model. All the signals are
truncated at the failure threshold, which is $D=1000$ for both
clusters. We evaluate the performance of our methodology under complete
as well as sparse scenarios. For a complete signal, the measurement
time points are preset at an equally spaced grid ${c_0,\ldots, c_{80}}$
on $[0,20]$ with $c_0=0, c_{80}=20$. For a sparse signal, we uniformly
sample $12$ time points from ${c_0,\ldots, c_{80}}$. Note that these
time points are prespecified; due to truncation, degradation signals at
these time points are not always observable. In this simulation, we
have around $40$ observations per complete signal and only around $6$
observations per sparse signal. Examples of the generated complete and
sparse degradation signals can be found in Figure~\ref{figsimusignals}.
In both plots, the black lines/dots represent the signals from cluster
$1$ and the grey lines/dots represent the signals from cluster~$2$.\looseness=1

\subsection{Estimation}\label{study115.5}

\begin{figure}

\includegraphics{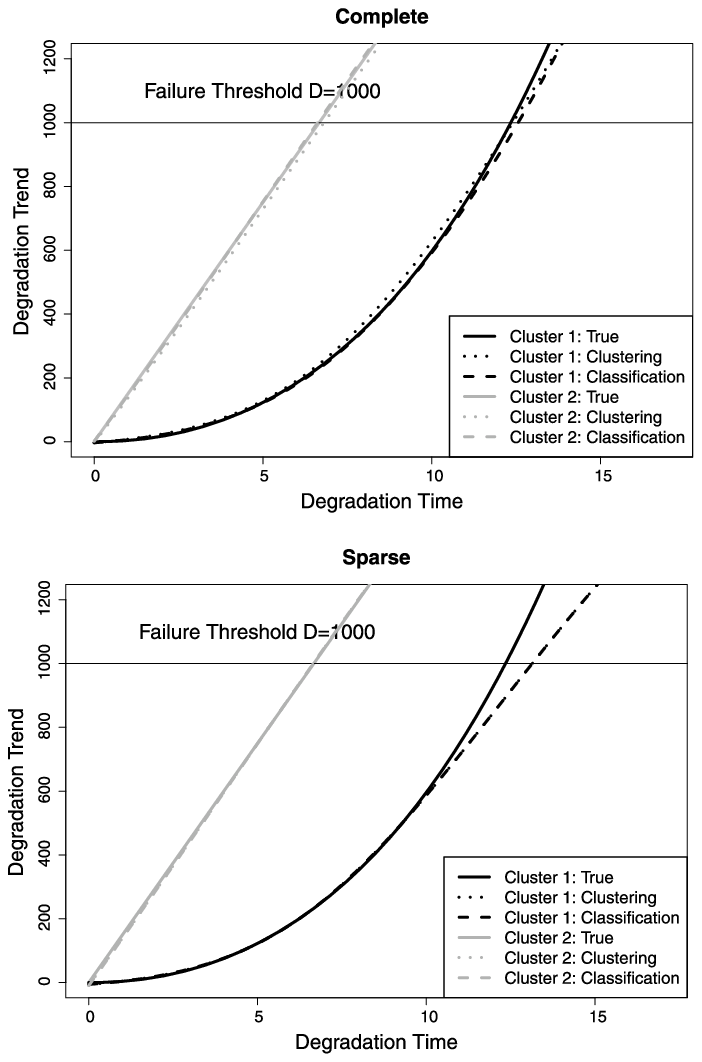}

\caption{Estimated (true, classification, clustering) mean function for
the complete and sparse scenarios, respectively.}\label{figsimumean}
\end{figure}

Based on the generated complete or sparse signals, we can estimate the
mean degradation trend for each cluster. The results are shown in the
top and bottom plots of Figure~\ref{figsimumean}. In both plots, the
solid, dotted and dashed lines represent the true mean trend, the
estimated trend in the clustering scenario and the estimated trend in
the classification scenario, respectively. From Figure~\ref
{figsimumean}, we observe that the estimated mean degradation trend for
both classification and clustering scenarios are very close to the true
trend (except for a small departure in the first cluster under the
sparse case, which is mainly due to the limited data available in that
region). This indicates that the mean functions are estimated
accurately for both classification and clustering scenarios and under
both complete and sparse cases.

In the clustering scenario, we are also interested in whether the
training signals are clustered accurately. To this purpose, we compute
the Rand index measuring the percentage of pairs of components on which
two clusterings, denoted by $X_1$ and $X_2$, agree or disagree [\citet{RandInd}]. Generally, $\operatorname{Rand}({X_1},{X_2})$ ranges from $0$ when there are
no pairs classified in the same class under $X_1$ and $X_2$, to $1$
when $X_1$ and $X_2$ give identical clustering. Here we use $X_1$ to
denote the true cluster membership of the 100 training signals
generated in one run of simulation and $X_2$ to denote the grouping
estimated using our proposed clustering method. Under both complete and
sparse scenarios, $\operatorname{Rand}({X_1},{X_2})=1$, which indicates that our
clustering method performs well under both scenarios for the generated
100 training signals. Accurate grouping of signals results in a better
prediction of RLD for fielded components.

The clustering performance is dependent upon many factors, for
instance, the mixing level of signals from different clusters. For
illustrative purposes, we perform a sensitivity analysis provided in
the supplemental material [\citet{ZSG14}].

\subsection{Prediction}\label{study12}
\begin{figure}

\includegraphics{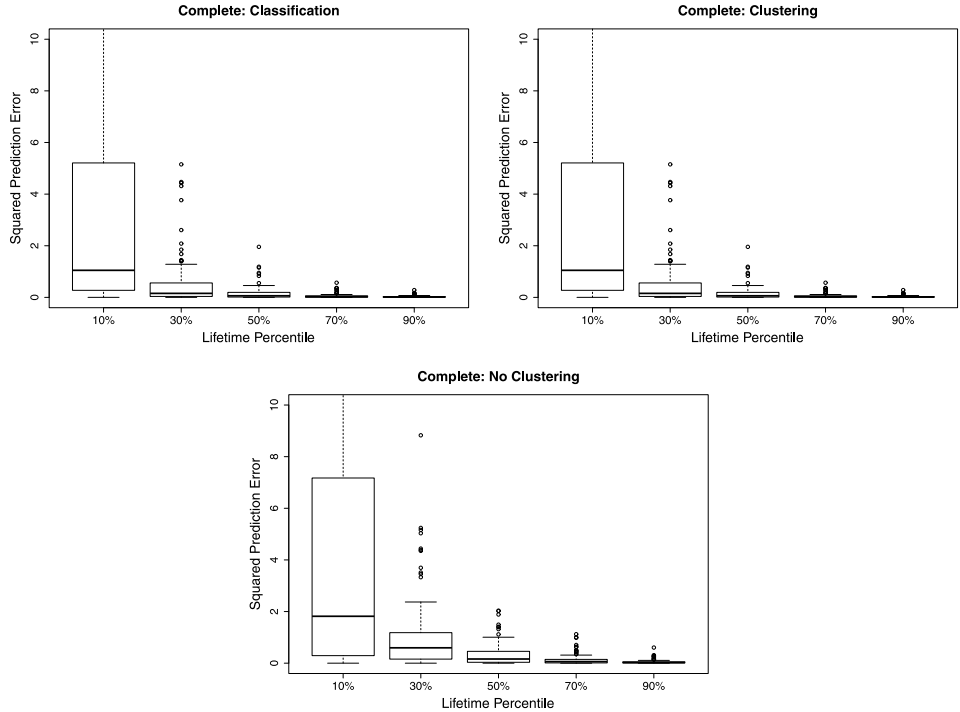}

\caption{Mean squared prediction errors for ``classification,''
``clustering'' and ``no clustering'' based on complete degradation signals.}\label{figsimuerrorcomplete}
\end{figure}

\begin{figure}[t]

\includegraphics{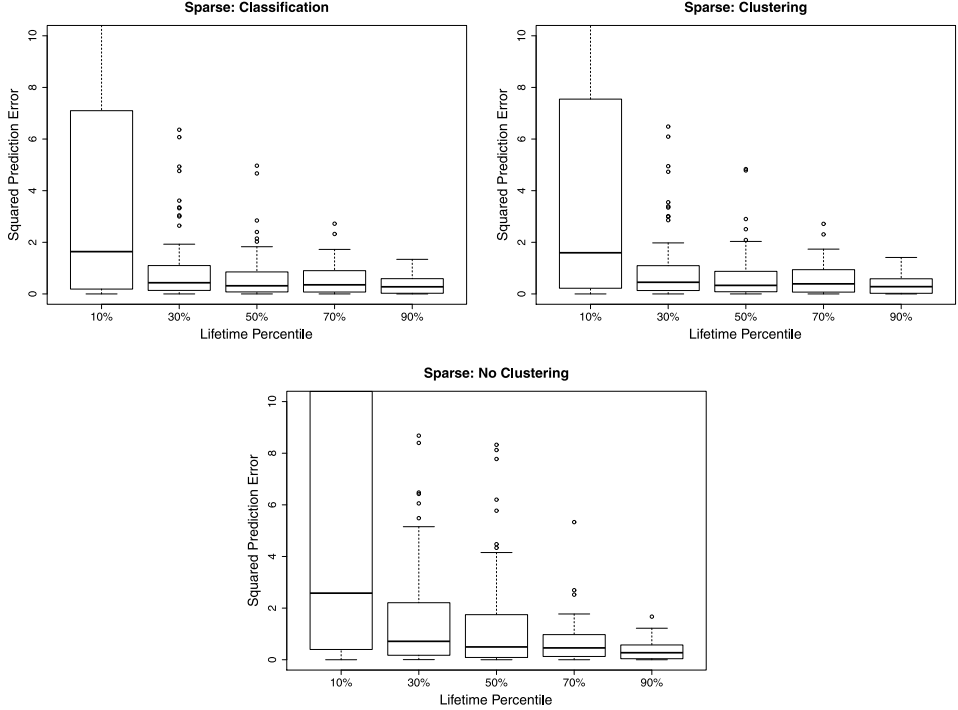}

\caption{Mean squared prediction errors for ``classification,''
``clustering'' and ``no clustering'' based on sparse degradation signals.}\label{figsimuerrorsparse}
\end{figure}

Our next step is to evaluate the performance of our model in terms of
residual life prediction. To assess the prediction accuracy, we use the
mean squared prediction error criteria because the posterior mean
(i.e., the expectation of the posterior predictive distribution) is
used as the point prediction. For each testing component, we compute
the prediction errors at the following percentiles of its entire life:
10\%, 30\%$, \ldots, 90$\% (90\% implies that 90\% of the component's
life has passed). The results based on complete and sparse signals are
illustrated in Figures~\ref{figsimuerrorcomplete}~and~\ref{figsimuerrorsparse}, respectively. In both figures, the top left and
top right plots are for the classification and clustering cases. For
comparative purposes, we also use a benchmark method, which is based on
our proposed framework (including the estimation and prediction
approaches). In the benchmark method all the components are assumed to
come from the same population and, therefore, it does not account for
the two different environmental types. We refer to this benchmark
method as ``no clustering.'' The results of ``no clustering'' are
reported in the bottom plots of Figures~\ref{figsimuerrorcomplete} and
\ref{figsimuerrorsparse}. For ease of comparison, we also summarize
the results in Table~\ref{tablepredictionerrors}, which gives the mean
and the variance of prediction errors for different methods based on
complete and sparse degradation signals.

One consistent observation from the figures and tables is that the
prediction results are very similar for the classification and
clustering scenarios. This again demonstrates that, at least in this
simulation, our proposed clustering algorithm can accurately classify
signals of similar patterns into the same group and separate signals of
distinct patterns into different groups. Furthermore, the benchmark
method provides less accurate predictions of the residual life of
components operated in the field. This is because the assumption that
all components are from the same population does not hold in this
simulation. The difference in performance between our methods and the
benchmark method ``no clustering'' is more significant at smaller life
percentiles, when the prior distribution plays a relatively more
important role in the RLD predictions.

\begin{table}[b]
\caption{Lifetime prediction results of ``classification,''
``clustering'' and ``no clustering'' under complete and sparse scenarios}\label{tablepredictionerrors}
\tabcolsep=0pt
\begin{tabular*}{\tablewidth}{@{\extracolsep{\fill}}@{}lccccc@{}}
\hline
\textbf{Lifetime percentiles} & \textbf{10\%} & \textbf{30\%} & \textbf{50\%} & \textbf{70\%} & \textbf{90\%}\\
\hline
Complete: classification & 3.24 & 0.58 & 0.21 & 0.10 & 0.06 \\
Complete: clustering & 3.24 & 0.58 & 0.21 & 0.10 & 0.06 \\
Complete: no clustering & 4.58 & 1.08 & 0.45 & 0.18 & 0.08 \\[3pt]
Sparse: classification & 6.18 & 1.09 & 0.62 & 0.53 & 0.34 \\
Sparse: clustering & 6.41 & 1.11 & 0.64 & 0.54 & 0.34 \\
Sparse: no clustering & 7.52 & 1.67 & 1.33 & 0.65 & 0.33 \\
\hline
\end{tabular*}
\end{table}

\section{Bearing case study}\label{study2}

\begin{figure}[t]

\includegraphics{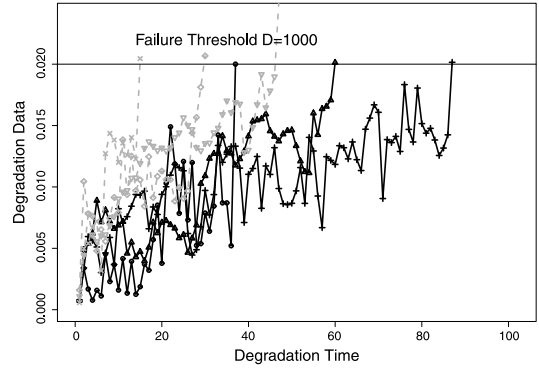}

\caption{Examples of bearing degradation signals.}\label{figbearingsignals}
\end{figure}

Bearings play an important role in a wide range of engineering
applications, particularly in rotating machinery. Failures of bearings
can lead to unexpected shutdown or failure of the entire engineering
system. In this study, we conduct an experiment to monitor the
degradation processes of rolling bearings. Each bearing is operated
under one of the following two rotational levels: $2200$ r.p.m. and $2600$
r.p.m. (r.p.m. is shorted for ``revolutions per minute''). The sample size of
each cluster is $16$ and $18$, respectively. For all bearings, we
collect vibration-based degradation signals up to their failures. The
failure threshold is prespecified as $D=0.02$ v.r.m.s. (v.r.m.s. is shorted for
``vibrational root mean square''). Examples of the resulting
degradation signals are in Figure~\ref{figbearingsignals}. In this
figure, the solid lines represent the degradation signals from cluster
1 ($2200$ r.p.m.) and the dashed lines represent those from cluster 2
($2600$ r.p.m.).

\begin{figure}[b]

\includegraphics{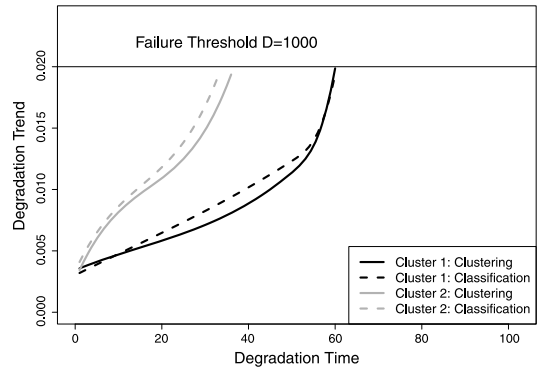}

\caption{Estimated mean degradation trend under the classification and clustering scenarios.}\label{figbearingmean}
\end{figure}


To evaluate the performance of our proposed degradation model, we
repeat the following study for $50$ times. Each time we randomly select
$5$ signals from each cluster as the testing signals. For these testing
signals, we assume that their cluster membership, or rotational speed,
is unknown and needs to be predicted. They are used to assess the
prediction performance. The rest of the $24$ degradation signals ($11$
of them are from cluster 1 and the rest of the $ 13$ signals are from
cluster~2) form a historical data set and they are used to train the
proposed degradation model and estimate the parameters. Depending on
the scenario we are interested in, whether it is ``classification'' or
``clustering,'' the cluster membership of the training components may
or may not be known. In Figure~\ref{figbearingmean}, we show the
estimated mean \mbox{degradation} trend (up to the failure threshold) for both
clusters. Apparently, the degradation processes in cluster 2 with a
rotational speed of $2600$ r.p.m. are relatively faster than those from
cluster 1 with a rotational speed of $2200$ r.p.m. Another observation is
that the estimated mean degradation trend under the classification and
clustering scenarios is similar.

\begin{figure}[b]

\includegraphics{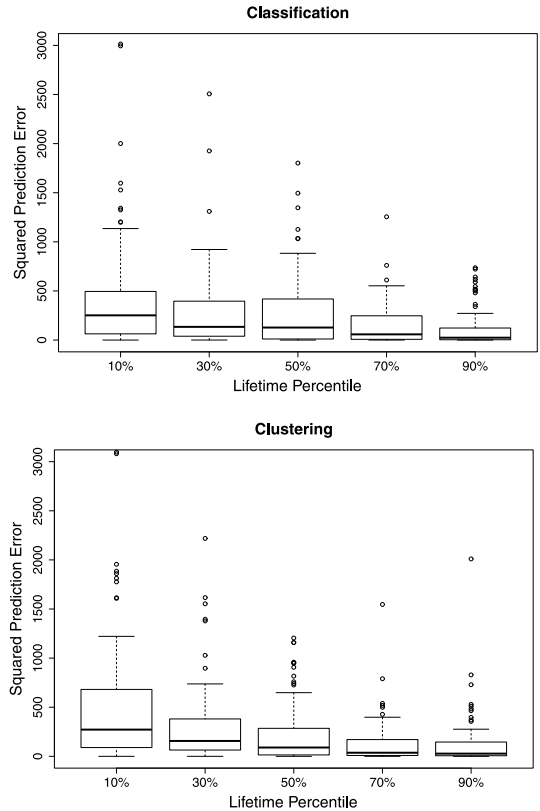}

\caption{Prediction errors under the classification and clustering scenarios.}\label{figbearingprediction}
\end{figure}

To mimic and illustrate the real-time updating process, we also assess
the prediction performance progressively. More specifically, for each
test bearing, we predict its residual life by using the partially
observed signal at the following percentiles of its lifetime: $10\%$,
$30\%,\ldots, 90\%$. As the percentile gets larger, we have more
degradation observations available and, therefore, we expect to see
more accurate and more precise predictions of the RLD. This is
demonstrated in the boxplots of Figure~\ref{figbearingprediction}, in
which we consider both the classification and clustering scenarios. In
these boxplots, the $x$-axis represents the lifetime percentiles and the
$y$-axis records the mean squared prediction errors. We observe that both
the median and variance of the prediction errors decrease as the
lifetime percentile increases, and this is consistent with our
observations from the simulation study. Another observation from
Figure~\ref{figbearingprediction} is that the prediction performance
for the classification and clustering scenarios is very similar. This,
once again, demonstrates that our proposed clustering algorithm can
classify degradation signals quite accurately.

\section{Summary}\label{discuss}

In this paper we propose a nonparametric model for characterizing the
evolution of degradation signals under varying experimental or
environmental conditions. This model can be used for predicting the
lifetime or residual life distributions of engineering components that
are still operated in the field. Our proposed framework relies on a
series of assumptions as follows:

\begin{longlist}[(3)]
\item[(1)] The underlying degradation process is smooth.

\item[(2)] The degradation signals follow a Gaussian process with
nonparametric mean and covariance.

\item[(3)] The environmental conditions can be categorized into a discrete
number of groups.

\item[(4)] The environmental conditions are constant over time.
\end{longlist}

In this paper we use the cubic B-spline basis due to its flexibility.
Other choices of basis functions can also be used depending on specific
assumptions on the smoothness of the degradation process. In our
simulation study, we observe that the estimation and prediction
performance of our model is robust to the departures from this
assumption (degradation signals from the first cluster cannot be
linearly expanded by the cubic B-spline basis functions). Nonetheless,
the use of cubic \mbox{B-}splines implies that the degradation is smooth over
time, which may not hold in all applications.

Our proposed model is nonparametric in the sense that the mean and
covariance of the Gaussian process specifying the conditional
distribution $S(\cdot)|Z$ are assumed not to have a predefined
parametric structure. This is a common approach in functional data
analysis. We have investigated the impact of departures from the
Gaussian assumption in a sensitivity study (not reported in the paper
but available from the authors). According to our sensitivity analysis,
the RLD prediction errors are more sensitive to the accuracy of the
estimated degradation trend functions compared to these distributional
assumptions.

The third assumption mentioned above may not always hold in real world
applications. In such cases, it may be more appropriate to consider the
environmental condition as a continuous covariate rather than discrete
clusters. We may still follow the general decomposition of the
degradation process in equation (\ref{model1}), but the classification or
clustering framework in equation (\ref{model2}) will not be applicable.
One possible approach is to follow similar ideas used in modeling the
ADT data, that is, by assuming certain functional, either linear or
nonlinear, relationships between the basis coefficient $\gamma_l$ and
the environmental variable $Z_l$.

In certain applications, the environmental conditions could be time
varying [\citet{Bian12}; \citet{geb08}]. For
instance, the cluster membership $Z_l$ may change at certain
deterministic or random time epochs. At these transitional epochs, the
observed degradation signals may be subject to sudden shocks, and the
rate at which the degradation progresses may also change. A~further
extension of the present framework to incorporate such time-varying
environmental conditions will be of interest in our future research.

\section*{Acknowledgements} We are thankful to the Editor and the
reviewers for their helpful comments and suggestions.

\begin{supplement}
\sname{Supplemental Meterial}
\stitle{Proofs and Derivations\\}
\slink[doi]{10.1214/14-AOAS749SUPP} 
\sdatatype{.pdf}
\sfilename{aoas749\_supp.pdf}
\sdescription{The supplemental material consists of two parts. In
Appendix~A, we present an available lemma that will be frequently used
in our \mbox{estimation} and prediction algorithms. In Appendix~B, we provide
details about our proposed EM algorithm for estimating the model parameters.}
\end{supplement}



\printaddresses
\end{document}